\pdfoutput=1
\documentclass{jpp}
\usepackage{hyperref}
\usepackage{graphics}
\usepackage{epstopdf, epsfig}

\newcommand{\apj}{ApJ}

\newcommand{\apjl}{ApJL}
\newcommand{\jgr}{J. Geophys. Res.}
\newcommand{\aap}{A~\&~A}

\newcommand{\mnras}{MNRAS}

\shortauthor{S. Yang, V.V. Pipin, D.D. Sokoloff, K.M. Kuzanyan and H. Zhang}

\title{The origin and effect of hemispheric helicity imbalance in solar dynamo}

\author{Shangbin YANG\aff{1,2}\corresp{\email{yangshb@nao.cas.cn}},
  V. V. Pipin\aff{3}, D. D. Sokoloff\aff{1,4,5}, K. M. Kuzanyan,\aff{1,5}
\and Hongqi Zhang\aff{1}}

\affiliation{\aff{1}Key Laboratory of Solar Activity, National
Astronomical Observatories, Chinese Academy of Sciences, 100101
Beijing, China
\aff{2}University of Chinese Academy of Sciences, 100049 Beijing, PR China
\aff{3}Institute of Solar-Terrestrial Physics, Russian Academy of
Sciences, Irkutsk, 664033, Russia
\aff{4}Department of Physics, Moscow University, 119992 Moscow, Russia
\aff{5}IZMIRAN,108840, Moscow, Russia}

\begin{document}

\maketitle
\begin{abstract}
In this paper we study the effects of hemispheric imbalance of magnetic helicity density on the hemispheric symmetry of
the dynamo generated large-scale magnetic field. Our study employs the axisymmetric dynamo model which takes into account the nonlinear effect of magnetic helicity conservation. 
We find that on the surface, the hemispheric imbalance of the magnetic helicity follows the evolution of the parity of the large-scale magnetic field. Random fluctuations of the $\alpha$-effect and the helicity fluxes can inverse the causal relationship, i.e., 
the magnetic helicity imbalance or the imbalance of magnetic helicity fluxes can drive the magnetic parity breaking. We also found that evolution of the net magnetic helicity of the small-scale fields follows the evolution of the net magnetic helicity of the large-scale fields with some time lag. We interpret this as an effect of the difference 
of the magnetic helicity fluxes out of the Sun from the large and
small scales. 
\end{abstract}

\section{Introduction}

The reflection asymmetry of the solar magnetic activity about equator
is one of the most important property of the solar dynamo. The magnetic
fields of the leading and following sunspots' groups of the solar
bipolar regions have predominantly opposite polarities in each hemisphere.
This is the so-called Hale polarity rule. The similar asymmetry exist
for the polar magnetic fields, which is the most prominent during
the sunspot minims. After \citet{Parker1955} , it is commonly accepted
that the reflection properties of the large-scale magnetic field are
determined by the dynamo mechanism operating inside the Sun. The essential
parts of the large-scale dynamo are govern by the differential rotation
and the turbulent convective motions. In the convective zone of a
star the global rotation makes turbulent convective motions helical.
This results in the reflection asymmetry of the convective vortices
about equator and produces the dynamo generation $\alpha$-effect,
which transforms the global toroidal magnetic field to the poloidal
\citep{Parker1955,Krause1980}. The reflection hemispheric asymmetry
of the $\alpha$-effect results to the hemispheric asymmetry of the
helical properties of solar magnetic field. This phenomenon is called
the hemispheric helicity rule (hereafter HHR) and it is observed in
the number of the magnetic helicity tracers like the current helicity
in the solar active regions, chirality of the solar prominences etc.
{The standard HHR suggests the negative sign of the current helicity of solar ARs in the northern hemisphere and the positive in southern one. 
For the global magnetic field the opposite HHR is
expected \citep{Blackman2003}. In the ideal situation, there is a hemispheric balance of distributions of the current helicity density and the total magnetic helicity.}

{The origin of the HHR and its impact on the dynamo is extensively discussed in the literature (see, e.g., the recent review \citealt{Brandenburg2018}).
Recently \citet{Singh2018} found that in cycle 24 more than 20\%
of the vector magnetic field synoptic maps show violations of the
expected hemispheric sign rule. Reversals of the sign rule of the current helicity of solar active regions during the beginning and the end of cycles 22 and 23 have been reported by \citet{Zhang2010} and references therein. 
Similar reversals have been found at the end of cycle 24 for the magnetic helicity density by \citet{Pipin2019a}.
The origin of the HHR reversals was addressed in our previous paper using the mean-field dynamo models \citep{Pipin2013d}. It was found that the reversal of the sign of the small-scale magnetic helicity follows the dynamo wave propagating inside the convection zone and the spatial patterns of the magnetic helicity reversals reflect the processes which contribute to generation and evolution of the large-scale magnetic fields.}

{In the paper, the HHR will be characterized by the sign distribution and the hemispheric imbalance of the magnetic helicity parameters, such as the current and magnetic helicity densities. 
For the perfect HHR the imbalance is about zero and the sign rule is obeyed.} 
One of the reason of this imbalance could be the hemispheric imbalance of the magnetic helicity flux from the surface to the outer atmosphere.
The existence of the net helicity flux is still under debate. For
example, \citet{Georgoulis2009}, found that the helicity injection
through the solar photosphere associated with active region magnetic
fields was well-balanced over the solar cycle 23. On the other hand,
\citet{Yang2012} reported significant imbalance between helicity
fluxes of northern and southern hemispheres. Currently, it is unclear
to which extend the imbalance of helicity fluxes impacts the dynamo
processes inside the convection zone. It is also unclear how the imbalance
of helicity fluxes affect the hemispheric imbalance of the magnetic
helicity density. Another possible reason could be due to redistribution
of the magnetic helicity density over the spatial scale. Both effects
(helicity fluxes and helicity cascades) are governing by the complicated
magnetohydrodynamic processes which can easily destroy the equatorial
symmetry from time to time and produce the net magnetic helicity of
the Sun.

In the paper we model effects of magnetic helicity imbalance using
the mean-field magneto-hydrodynamic framework. In this case it is
important to distinguish magnetic helicity of the small-scale and
the large-scale (global) field of the Sun. We represent the magnetic
field $\mathbf{B}$ and its vector potential $\mathbf{A}$ ($\mathbf{B}=\nabla\times\mathbf{A}$)
to the sum of the mean and fluctuating parts: $\mathbf{B}=\overline{\mathbf{B}}+\mathbf{b}$,
$\mathbf{A}=\overline{\mathbf{A}}+\mathbf{a}$, where the overbar
denotes the mean quantities. Next, the magnetic helicity is defined
as integral over the closed domain $\mathcal{H}=\int\mathbf{A}\cdot\mathbf{B}dV$,
and the $\mathbf{A}\cdot\mathbf{B}$ is the magnetic helicity density.
Assuming the validity of the Reynolds rule for averaging of the products
and sum of the turbulent quantity we can distinguish between the contributions
of the large-scale and the small-scale magnetic field to the magnetic
helicity density: 
\begin{equation}
\overline{\chi}^{(tot)}=\overline{\mathbf{A}\cdot\mathbf{B}}=\overline{\mathbf{A}}\cdot\overline{\mathbf{B}}+\overline{\mathbf{a\cdot b}}.\label{chitot}
\end{equation}
Hereafter, we denote the small-scale and large-scale parts of the
magnetic helicity density as follows, $\overline{\chi}=\overline{\mathbf{a\cdot b}}$,
$\overline{\chi}^{\left(m\right)}=\overline{\mathbf{A}}\cdot\overline{\mathbf{B}}$.

Following to \citet{Hubbard2012,Pipin2013c}, we employ the conservation
law for $\overline{\chi}^{(tot)}$ :

\begin{equation}
\frac{d}{dt}\int\overline{\chi}^{(tot)}dV=-2\eta\int\left\{ \overline{\mathbf{B}}\cdot\mathbf{\overline{J}}+\overline{\mathbf{b\cdot j}}\right\} dV-\int\boldsymbol{\nabla\cdot}\boldsymbol{\boldsymbol{\mathcal{F}}}^{\chi}dV\label{eq:int-cons}
\end{equation}
where the $\boldsymbol{\boldsymbol{\mathcal{F}}}^{\chi}$ denotes
the helicity flux. 
{In the above cited papers it was shown, that with this formulation of the magnetic helicity conservation the dynamo evolution avoids the catastrophic quenching regimes.} 
The differential equation that corresponds to  Eq.(\ref{eq:int-cons}) is

\begin{equation}
\frac{\partial\overline{\chi}^{(tot)}}{\partial t}=-\frac{\overline{\chi}}{R_{m}\tau_{c}}-2\eta\overline{\mathbf{B}}\cdot\mathbf{\overline{J}}-\boldsymbol{\nabla\cdot}\boldsymbol{\boldsymbol{\mathcal{F}}}^{\chi}-\mathbf{\left(\overline{U}\cdot\boldsymbol{\nabla}\right)}\overline{\chi}^{(tot)}\label{eq:helcon}
\end{equation}
In deriving the the Eq.(\ref{eq:int-cons}) we assumed ${\displaystyle 2\eta\overline{\mathbf{b\cdot j}}=\frac{\overline{\chi}}{R_{m}\tau_{c}}}$
(see, \citealp{Kleeorin1999}), where the magnetic Reynolds number
$R_{m}=10^{3-6}$ and $\eta$ is the microscopic diffusivity. Note
that conservation law given by the Eq.(\ref{eq:int-cons}) take into
account the possibility of the magnetic helicity fluxes out of the
dynamo domain. In the stationary state we have locally:

\begin{equation}
\overline{\chi}\approx-\overline{\mathbf{A}}\cdot\overline{\mathbf{B}}=-\overline{\chi}^{\left(m\right)}\label{eq:lb}
\end{equation}
This balance can be changed in any direction by the helicity fluxes
either on the small or the large scales.

We assume that the magnetic helicity density balance is following
to the Eqs.(\ref{eq:helcon},\ref{eq:lb}). Clearly, there are important
unknown details in the the Eq.(\ref{eq:helcon}), in particular, those
are related to the helicity density fluxes. Further it will be shown
that breaking of the equatorial symmetry of the global magnetic field
can result in the hemispheric imbalance of the magnetic helicity density,
as well. We study the mutual effect this imbalance and the magnetic
parity breaking using mean-field dynamo models.

In Section 2 we describe some specific details of our dynamo model.
Section 3 is devoted to description of the main results and to discussion
of those results in the light of the available observational proxies.
Section 4 summarizes our findings.

\section{Basic equations}

\subsection{Dynamo model}

\begin{figure}
\includegraphics[width=0.95\textwidth]{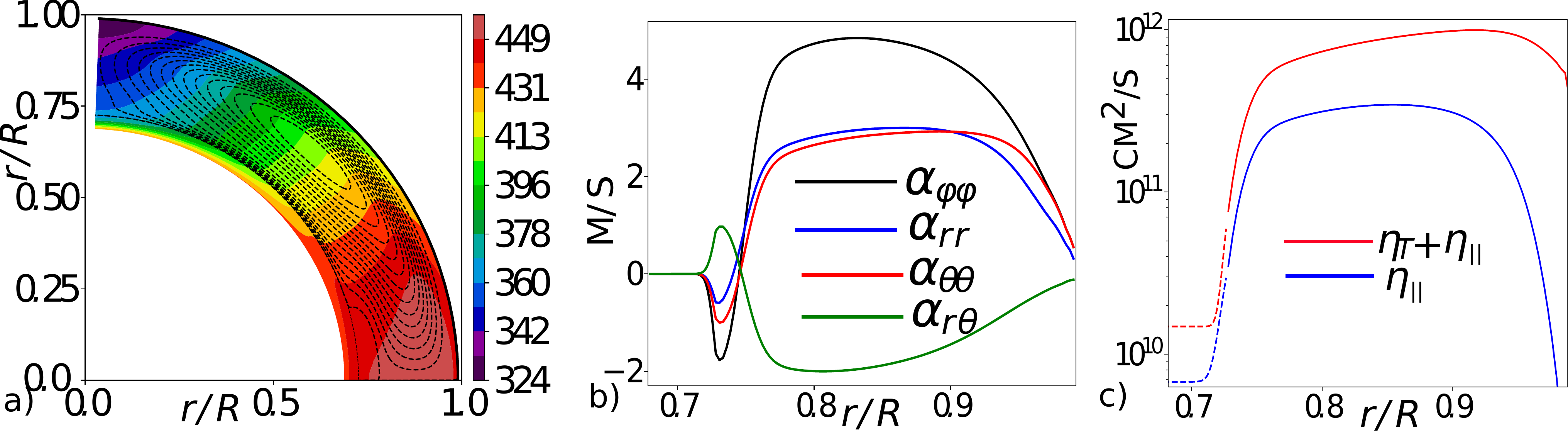} \caption{\label{fig1} a) The reference angular velocity and meridional circulation distributions; b) the radial profiles of the $\alpha$-effect tensor at latitude 45$^{\circ}$; c) radial profiles of the total, $\eta_{T}+\eta_{||}$,
and the rotationally induced part, $\eta_{||}$, of the eddy magnetic diffusivity.}
\end{figure}

In this paper we will discuss the kinematic version of the mean-field
dynamo model developed recently by \citealp{Pipin2019c,Pipin2018b}.
We study the mean-field induction equation in the turbulent  perfectly
conducting medium:

\begin{equation}
\frac{\partial\overline{\mathbf{B}}}{\partial t}=\boldsymbol{\nabla}\times\left(\boldsymbol{\mathcal{E}}+\overline{\mathbf{U}}\times\overline{\mathbf{B}}\right),\label{eq:dyn}
\end{equation}
where $\boldsymbol{\mathcal{E}}=\overline{\mathbf{u\times b}}$ is
the mean electromotive force, with $\mathbf{u,\,b}$ being fluctuating
velocity and magnetic field, respectively, $\overline{\mathbf{U}}$
is the mean velocity field, which is represented by the differential
rotation and meridional circulation. We assume that the large-scale
flow as well as the global thermodynamics of the convection zone remain
unaffected by the solar dynamo. Those effects were discussed in above
cited papers. A large-scale axisymmetric magnetic field is represented
by decomposition on the sum of the toroidal and poloidal parts: 
\[
\overline{\mathbf{B}}=\mathbf{e}_{\phi}B+\nabla\times\frac{A\mathbf{e}_{\phi}}{r\sin\theta},
\]
where $\theta$ is the polar angle. The mean electromotive force $\boldsymbol{\mathcal{E}}$
is expressed as follows: 
\begin{equation}
\mathcal{E}_{i}=\left(\alpha_{ij}+\gamma_{ij}^{(\Lambda)}\right)\overline{B}_{j}+\eta_{ijk}\nabla_{j}\overline{B}_{k}.\label{eq:EMF-1}
\end{equation}
The tensor $\alpha_{ij}$ represents the $\alpha$-effect, $\gamma_{ij}^{(\Lambda)}$
is the turbulent pumping, and $\eta_{ijk}$ is the diffusivity tensor.
The $\alpha$ effect includes hydrodynamic and magnetic helicity contributions,
\begin{eqnarray}
\alpha_{ij} & = & C_{\alpha}\left(1+\xi^{\left(\alpha\right)}\left(t,\theta\right)\right)\alpha_{ij}^{(H)}+\alpha_{ij}^{(M)},\label{alp2d}
\end{eqnarray}
{where $\xi^{\left(\alpha\right)}\left(t,\theta\right)$ is
the fluctuating part of the $\alpha$-effect. Further details about the kinetic part of the $\alpha$ effect, $\alpha_{ij}^{(H)}$, as well as $\gamma_{ij}^{(\Lambda)}$, $\alpha_{ij}^{(M)}$, and $\eta_{ijk}$
can be found in \citet{Pipin2019c}. The nonlinear feedback of the
large-scale magnetic field to the $\alpha$-effect is described by
a dynamical quenching due to the constraint of magnetic helicity conservation given by the Eq.(\ref{eq:helcon}). Similar to that paper the integration domain include the overshoot layer, which bottom is at $r_{b}=0.68R_{\odot}$.
The convection zone extends from $r_{b}=0.728R_{\odot}$ to $r_{e}=0.99R_{\odot}$.
We matched the dynamo solution to the potential field outside, and assume zero magnetic field at the bottom boundary. The numerical scheme employs the spatial mesh with 100 nodes in the radius. We use the pseudo-spectral approach for the differentiation operators along latitude
and the 64 nodes in latitude are located in the collocation points of the Legendre polynomial. 
Turbulent coefficients in the bulk of
convection zone are calculated from solution of the mean-field thermodynamic equation using the mixing-length approximation and the mean entropy distribution. 
It is assumed the in overshoot layer all the turbulent coefficients except the eddy diffusivity are exponentially quenched.
For the numerical stability we keep the finite eddy diffusivity in the overshoot layer. More details can be found in above cited paper of \citet{Pipin2019c}. Distributions of the angular velocity profile, meridional circulation, the kinetic part of the $\alpha$ effect tensor and the rotationally anisotropic eddy diffusivity in our model are shown in Figure \ref{fig1}. }

\subsection{Random sources of the helicity density imbalance}

\begin{figure}
\centering \includegraphics[width=0.8\textwidth]{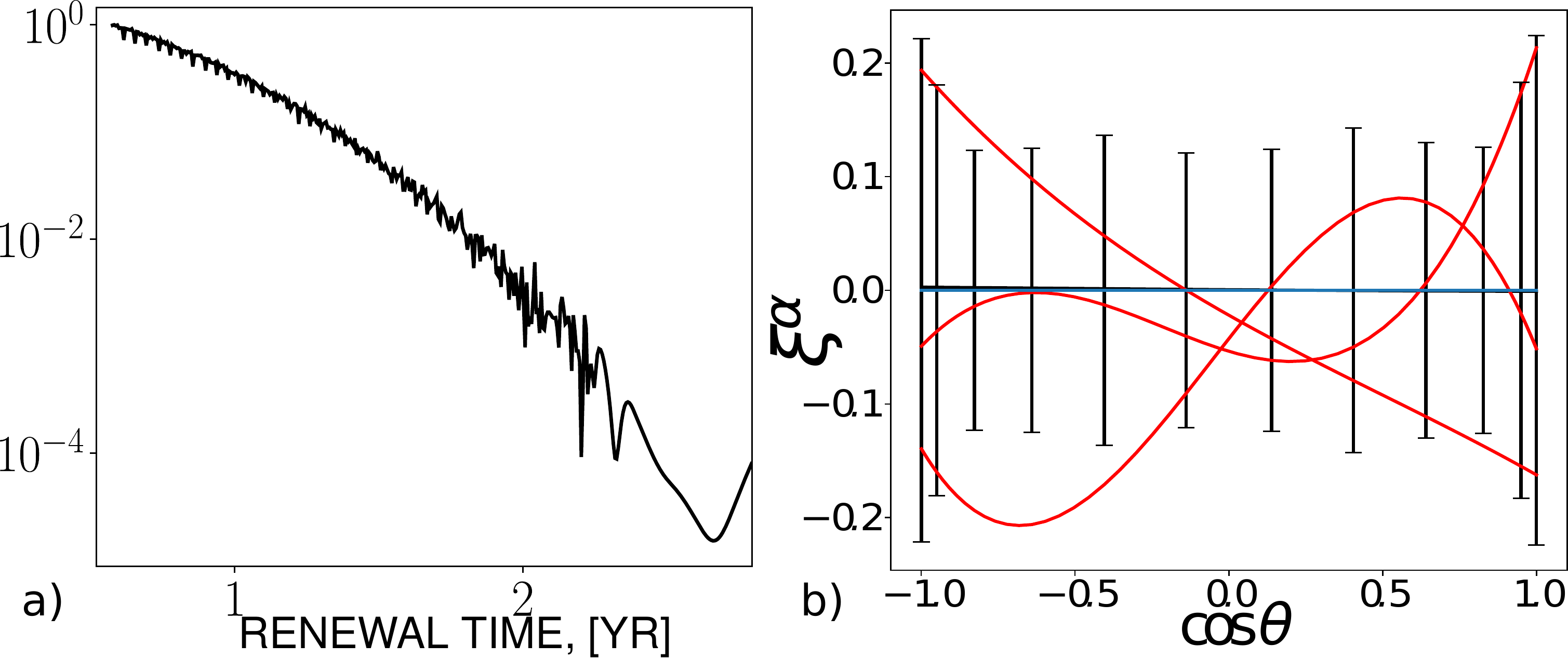}\caption{{\label{pdf}a) Probability distribution of the renewal time
intervals of $\xi^{\left(\alpha\right)}\left(t,\theta\right)$; b)
Red lines show three realizations of $\xi^{\left(\alpha\right)}\left(t,\theta\right)$,
the vertical bars show the standard deviations for particular latitudes
and the blue line shows the mean over ensemble of realizations, $\overline{\xi^{\left(\alpha\right)}\left(t,\theta\right)}\approx0$.}}
\label{fig:fluct} 
\end{figure}

In our model we explore a few possible sources of the helicity density
imbalance. The first is the non-symmetric about equator fluctuations
of the kinetic $\alpha$-effect (see the Eq.(\ref{alp2d} ). For the
deterministic problems like the dynamo equations system, the Eqs.(\ref{eq:helcon},\ref{eq:dyn})
which are solved by the standard numerical integration schemes, the
spatial and temporal fluctuations of the model parameters are the
sources of the potential numerical pitfalls because the meaning of
the derivative is rather different for deterministic and the random
functions. Practically, without going deep into details, we are safe
if the typical spatial and temporal scales of fluctuations are much
larger when the size of the spatial mesh and the size of the time-step.
To simulate the randomness of the $\alpha$-effect distribution over
hemispheres we generate the spatially random gaussian sequences, $\xi^{(\alpha)}\left(\theta_{j}\right)$
, where $\theta_{j}$ are the collocation points of the Legendre polynomials,
and $\overline{\xi^{(\alpha)}\left(\theta_{j}\right)}=0$, $\sigma\left(\xi^{(\alpha)}\left(\theta_{j}\right)\right)=0.5$.
Then we decompose the sequence $\xi^{(\alpha)}\left(\theta_{j}\right)$
in the Legendre polynomials and filter out all the Legendre modes
higher than $\ell=5$. The resulted latitudinal fluctuations of the
$\alpha$-effect are described via the smooth functions. The ensemble
of the $\xi^{(\alpha)}\left(\theta\right)$ follows the Gaussian probability
distribution with mean approximately equals 0 and the the standard
deviation $\sigma\approx0.2$. The renewal time for the sequences
$\xi^{(\alpha)}$ is also taken in form of the random sequence. From
that we pick up the values larger than 0.3Yr intervals, which is safe
for the numerical scheme with the time step about few hours. The probability
distribution of the renewal time is shown in the Figure \ref{fig:fluct}a.
{
The fluctuations of the $\alpha$-effect in latitude are illustrated in Figure \ref{fig:fluct}b. Note that contribution of the $\xi^{\left(\alpha\right)}\left(t,\theta\right)$
in the Eq.(\ref{alp2d}) is multiplied by factor $\cos\theta$ caused by the $\alpha_{ij}^{(H)}$.}

{Another source of the hemispheric imbalance of the helicity density can be due to the asymmetry about the equator of the flux of helicity density from the dynamo domain to the corona.} 
Following the suggestions
by \citep{Guerrero2010} we model this by subtracting the fraction
of the helicity density from the local helicity density in the upper
parts of the convection zone. Thus, the modified equation for the
helicity density evolution is

\begin{eqnarray}
\frac{\partial\overline{\chi}^{(tot)}}{\partial t} & = & -\frac{\overline{\chi}}{R_{m}\tau_{c}}-2\eta\overline{\mathbf{B}}\cdot\mathbf{\overline{J}}-\boldsymbol{\nabla\cdot}\boldsymbol{F}\label{eq:helcon-1}\\
 &  & -\frac{\tau_{\xi}\left(r\right)}{\tau_{0r}}\sin^{2}\theta\left(\xi^{(\chi)}\left(t,\theta\right)\overline{\chi}+\xi^{(m)}\left(t,\theta\right)\overline{\chi}^{(m)}\right),\nonumber 
\end{eqnarray}
where, $\boldsymbol{\boldsymbol{F}}=-\eta_{\chi}\boldsymbol{\nabla}\overline{\chi}$,
with $\eta_{\chi}=0.1\eta^{(I)}$. 
{Similarly to \citet{Pipin2013d},
we employ $R_{m}=10^{6}$. The last term in the Eq.(\ref{eq:helcon-1})
takes into account the helicity density flux out of the solar convection zone. 
In the paper we study the fluxes of the small-scale and the
large-scale magnetic helicities. They are $\xi^{(\chi)}\overline{\chi}$
and $\xi^{(m)}\overline{\chi}^{(m)}$, respectively. It is assumed
that the fluxes are due to the near surface magnetic activity. Therefore, we apply the factor $\sin^{2}\theta$ and introduce the function:
\begin{equation}
\tau_{\xi}\left(r\right)=\frac{1}{2}\left[1-\mathrm{erf}\left(100\left(r_{0}-r\right)\right)\right]\,,\label{eq:tau}
\end{equation}
where $r_{0}=0.9R_{\odot}$ and the dimension factor corresponds to
the maximum of the PDF of the renewal time, $\tau_{0r}=0.5$Yr (see,
Figure \ref{pdf}a).}

The sequence of the renewal times of the helicity density outflows
will be determined in the same way as for the $\alpha$-effect except
for the low limit which is put about ten times smaller and it is equal
to one month. Thus, the net helicity density flux over hemisphere
is computed as integral over the shell which includes subsurface region
between $r_{0}$ and $R_{\odot}$. Functions $\xi^{(\chi)}$ and $\xi^{(m)}$
are random in latitude, and they are defined in the same way as $\xi^{(\alpha)}$
and we use the ensembles of the spatial fluctuations with the Gaussian
probability function distribution, the mean value of $\overline{\xi^{(\chi,m)}}=0$
and the standard deviation of $\sigma\left(\xi^{(\chi,m)}\right)=1$.
These fluctuations are driven with the random renewal time interval, which has the same probability distribution as the $\alpha$ effect fluctuations. 
{For the magnetic helicity density we employ
$\overline{\chi}=0$ at the bottom and $\nabla_{r}\overline{\chi}=0$
at the top of the convection zone domain. 
We neglect the magnetic helicity evolution in the overshoot region for the sake of simplicity.}

{For the purpose of analysis we introduce the total energy
of the dynamo generated magnetic field,
\begin{equation}
E_{B}=\frac{1}{8\pi}\int\boldsymbol{\overline{B}}^{2}\mathrm{dV},\label{eq:en}
\end{equation}
where integration is over the dynamo domain. Following \citet{Pipin2012},
we mimic the sunspot number using parameters of the toroidal magnetic field in subsurface shear layer: 
\begin{equation}
W=\overline{B}_{\mathrm{max}}\left(t\right)\exp{\displaystyle \left(-\frac{\overline{B}_{\mathrm{max}}\left(t\right)}{B_{0}}\right)}\label{eq:sn}
\end{equation}
where $B_{0}=600$G and $\overline{B}_{\mathrm{max}}\left(t\right)=\max_{\mu=-1:1}\left(\left(\overline{B}\left(\mu,t\right)\right|_{0.95R}\right)$. In the Eq(\ref{eq-sn}), we assume that sunspots are produced from the toroidal magnetic fields by means of some nonlinear instability.}

The properties of the equatorial symmetry of the magnetic activity is characterized by the parity index. Let's define the parameters characterizing the energy of the symmetric and antisymmetric parts of the radial magnetic field at the surface

\begin{eqnarray*}
\overline{E}_{B}^{S} & =\frac{1}{4} & \int_{-1}^{1}\left[\overline{B}_{r}\left(\mu,t\right)+\overline{B}_{r}\left(-\mu,t\right)\right]^{2}d\mu,\\
\overline{E}_{B}^{N} & =\frac{1}{4} & \int_{-1}^{1}\left[\overline{B}_{r}\left(\mu,t\right)-\overline{B}_{r}\left(-\mu,t\right)\right]^{2}d\mu.
\end{eqnarray*}
 Then the parity index, or the reflection symmetry index for this
component of the magnetic activity is

\begin{equation}
P=\frac{\overline{E}_{B}^{S}-\overline{E}_{B}^{N}}{\overline{E}_{B}^{S}+\overline{E}_{B}^{N}}.\label{eq:parity}
\end{equation}
 {Also, we define the parameters characterizing the helicity fluxes of the large- and small-scale magnetic field. In the model, these fluxes are determined by the random parameters $\xi^{(\chi)}$ and
$\xi^{(m)}$. For the large-scale magnetic field we define the latitudinal
helicity density flux: 
\begin{equation}
F_{L}=-2\pi\sin^{2}\theta\xi^{(m)}\left(t,\theta\right)\int_{.9R}^{R}\frac{\tau_{\xi}\left(r\right)}{\tau_{0r}}\overline{\chi}^{(m)}r^{2}\mathrm{dr}.\label{eq:fL}
\end{equation}
The small-scale magnetic helicity density flux includes the the diffusive
flux, as well, see the Eq.(\ref{eq:helcon-1}):
\begin{eqnarray}
F_{S} & = & \left.-2\pi R^{2}\eta_{\chi}\boldsymbol{\nabla}\overline{\chi}\right|_{.9R}^{R}-2\pi\sin^{2}\theta\xi^{(\chi)}\left(t,\theta\right)\int_{.9R}^{R}\frac{\tau_{\xi}\left(r\right)}{\tau_{0r}}\overline{\chi}r^{2}\mathrm{dr}.\label{eq:fl}
\end{eqnarray}
Following \citet{Berger2000,Hawkes2019AA} we define the latitudinal
helicity density flux due to the differential rotation:
\begin{equation}
F_{\Omega}=-4\pi R^{3}\sin\theta A\overline{B_{r}}\ \overline{U}_{\phi},\label{eq:dr}
\end{equation}
where $A$ is the vector-potential and $\overline{U}_{\phi}=R\sin\theta\Omega$$\left(R,\theta\right)$
is the large-scale azimuthal flow at the surface. 
This helicity flux does not enter directly in the helicity evolution equation. Its effect on the dynamo is determined by the boundary conditions.}

\begin{figure}
\begin{centering}
\includegraphics[width=0.8\textwidth]{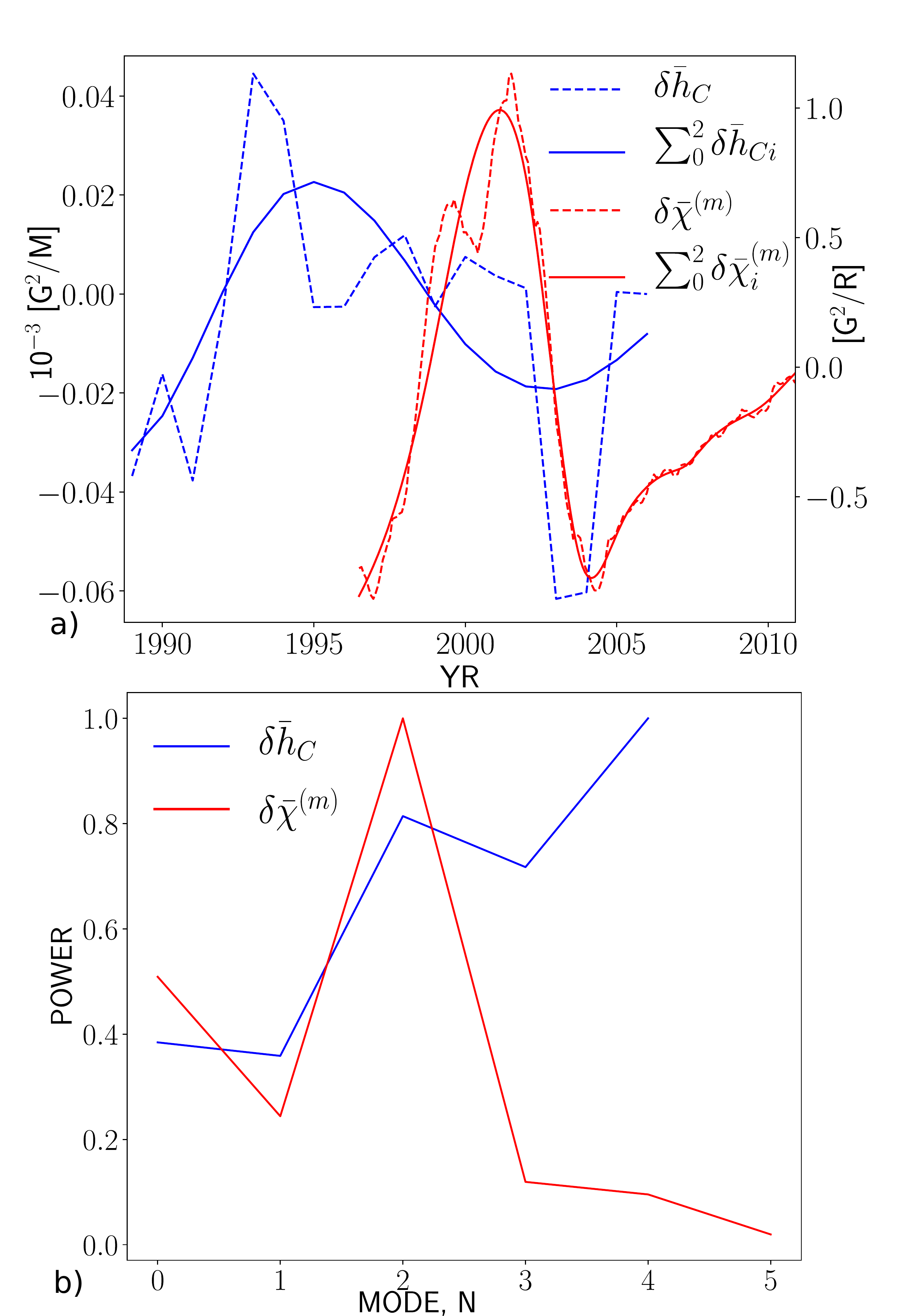} 
\par\end{centering}
\caption{(a) {Current helicity density imbalance of the solar ARs (blue color) and the large-scales magnetic field helicity density imbalance
(normalized to solar radius, shown in red color), and their representation via the first three empirical modes;} (b) shows the relative power of each mode in the empirical modes decomposition (EMD), where the results are normalized to the maximum of the magnitude of the signal.}
\label{fig:imb} 
\end{figure}

\section{Observational proxies of the magnetic helicity imbalance}

{In the Introduction we defined the hemispheric helicity rule by the surface integral of the magnetic helicity proxies.
} 
Figure \ref{fig:imb}(a) shows the integral of the current helicity density of the solar active regions obtained from the reduced data set of Huairou Solar Observing Station given in \citet{Zhang2010},
\begin{equation}
\delta\overline{h}_{C}=\int_{0}^{1}\overline{h}_{C}d\mu\,
\end{equation}
{where $\mu=\cos\theta$ and $\theta$ is the polar angle},
and the same for the magnetic helicity density of the global magnetic field which was reconstructed by Pipin and Pevtsov \citeyearpar{Pipin2014b}
using the SOHO/MDI data set, 
\begin{equation}
\delta\overline{\chi}^{\left(m\right)}=\int_{0}^{1}\overline{\chi}^{\left(m\right)}d\mu\,
\end{equation}
{We observe the solar cycle variations of the HHR parameters in both cases. The low cadence of data set of $\delta\overline{h}_{C}$
and the limited time interval in both data sets result into uncertainty in our conclusions about the long-term behavior of these parameters.
To get a rough idea we apply the empirical mode decomposition (EMD) method.
Because of the mentioned issue of our data sets, our analysis is rather rough and it can be subjected to systematic aliasing errors.} We show results in the same Figure \ref{fig:imb}(a). {The information about the contribution of the empirical modes to the energy of the signal is shown in Figure \ref{fig:imb}(b). In the signal of
the $\delta\overline{h}_{C}$, the "small-scale" modes of short
periods 1-3 year are the strongest. Their effect on the whole $\delta\overline{h}_{C}$
is rather strong over the sunspot minima. The first three modes of $\delta\overline{h}_{C}$ show the variation with the solar cycle period.} 
The large-scale magnetic helicity density imbalance has a strong signal with period of about 9 years and the first three modes
quite accurately reproduce the total signal. The sum of the first
three modes of current helicity density imbalance. $\sum_{0}^{2}\delta\overline{h}_{C\,i}$
has a similar period. It is seen that the $\sum_{0}^{2}\delta\overline{h}_{C\,i}$
goes ahead of the $\sum_{0}^{2}\delta\overline{\chi}_{i}^{\left(m\right)}$
with the phase shift about $\pi$. {This rough analysis shows a possibility of the quasi-regular variations of $\delta\overline{h}_{C}$
and $\delta\overline{\chi}^{\left(m\right)}$ in the dynamo cycle.
We shall see whether this effect can be reproduced in our dynamo models.}

\section{Results}

{The dynamo model governing parameters are the same as in the paper of \citet{Pipin2019c}.} 
{In all the runs we consider
a slightly overcritical dynamo regimes using the same dynamo parameters set as in the our previous papers . 
Similar to those papers our models are weakly nonlinear with $\beta_{max}=\left|B\right|/\sqrt{4\pi\bar{\rho}u'^{2}}<0.4$.
}
The random parameters in the models are applied in following the
Table~\ref{Tab1}. {Also, the Table shows some output parameters, like the amplitude of the helicity fluxes variations, the amplitude of the total magnetic energy and amplitudes of helicity imbalances in our models. }

\begin{table}
\centering{}%
\begin{tabular}{lcccp{1.2cm}p{1.2cm}p{1.2cm}p{1.2cm}p{1.2cm}p{1.2cm}}
Model  & $\xi^{(\alpha)}$  & $\xi^{(\chi)}$  & $\xi^{(m)}$  & \noindent {$F_{L}$,}10$^{40}$

\noindent {{[}Mx$^{2}$/d{]}} & \noindent {$F_{S}$,}10$^{42}$

\noindent {{[}Mx$^{2}$/d{]}} & \noindent {$F_{\Omega}$,}10$^{43}$

\noindent {{[}Mx$^{2}$/d{]}} & \noindent {$E_{B}$,}10$^{36}${ {[}G$^{2}${]}} & \noindent {$\delta\overline{\chi}$,}10$^{12}$

\noindent {{[}G$^{2}$M{]}} & \noindent {$\delta\overline{\chi}^{\left(m\right)}$,}10$^{12}$

\noindent {{[}G$^{2}$M{]}}\tabularnewline
M1  & yes  & no  & no  & \noindent - & \noindent 7  & \noindent 1.9  & \noindent 4.3 & \noindent 8.4 & \noindent 14.2\tabularnewline
M2  & $\overline{\xi^{(\alpha)}}$  & no  & yes  & \noindent - & \noindent 8.8 & \noindent 2.7 & \noindent 5.1 & \noindent 7.3 & \noindent 9.5\tabularnewline
M3  & $\overline{\xi^{(\alpha)}}$  & yes  & no  & \noindent 0.8 & \noindent 7 & \noindent 1.5 & \noindent 4.5 & \noindent 6 & \noindent 7.5\tabularnewline
M4  & $\overline{\xi^{(\alpha)}}$  & yes  & yes & \noindent 0.7 & \noindent 7 & \noindent 2 & \noindent 4.5 & \noindent 4.1 & \noindent 7.1\tabularnewline
\end{tabular}\caption{\label{Tab1}Parameters of the model runs. In the models M2, M3 and
M4 we neglect the hemispheric asymmetry of the $\alpha$-effect fluctuations.
In this case the fluctuating part of the $\alpha$ -effect is equal
to $\overline{\xi^{(\alpha)}}$$\left(t\right)$ where we use average
over latitudes{, }see the Eq(\ref{alp2d}). The forth column
shows the maximum of the total magnetic energy in the dynamo domain;
the imbalances}
\label{tab:kd} 
\end{table}

\begin{figure}
\centering \includegraphics[width=0.8\textwidth]{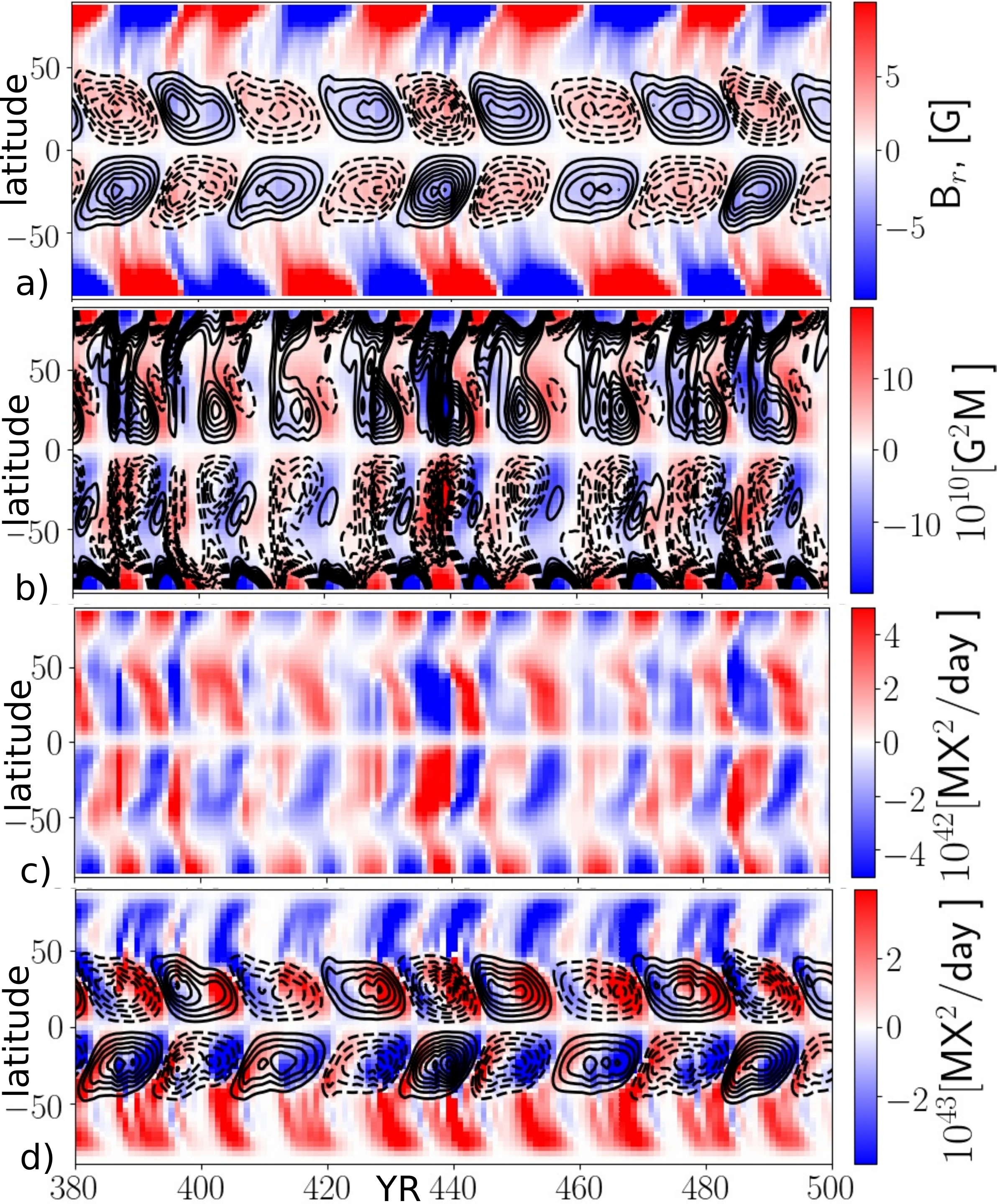}\caption{\label{fig-tl}The model M2: a) the time-latitude diagram of the toroidal magnetic field in the subsurface shear layer (contours in range of $\pm600$G) and the radial magnetic field at the surface (color image);
b) the time-latitude evolution of the small-scale (color image) and the large-scale magnetic helicity density (contours are drawn for the the same range of magnitudes $\pm20$ $10^{10}$ G$^{2}$M); c) the small-scale magnetic helicity density flux, $F_{S};$ 
d) the time-latitude diagram of the toroidal magnetic field (same as the panel (a) and the
magnetic helicity density flux from differential rotation, $F_{\Omega}$(color image).}
 
\end{figure}

\begin{figure}
\centering \includegraphics[width=0.8\textwidth]{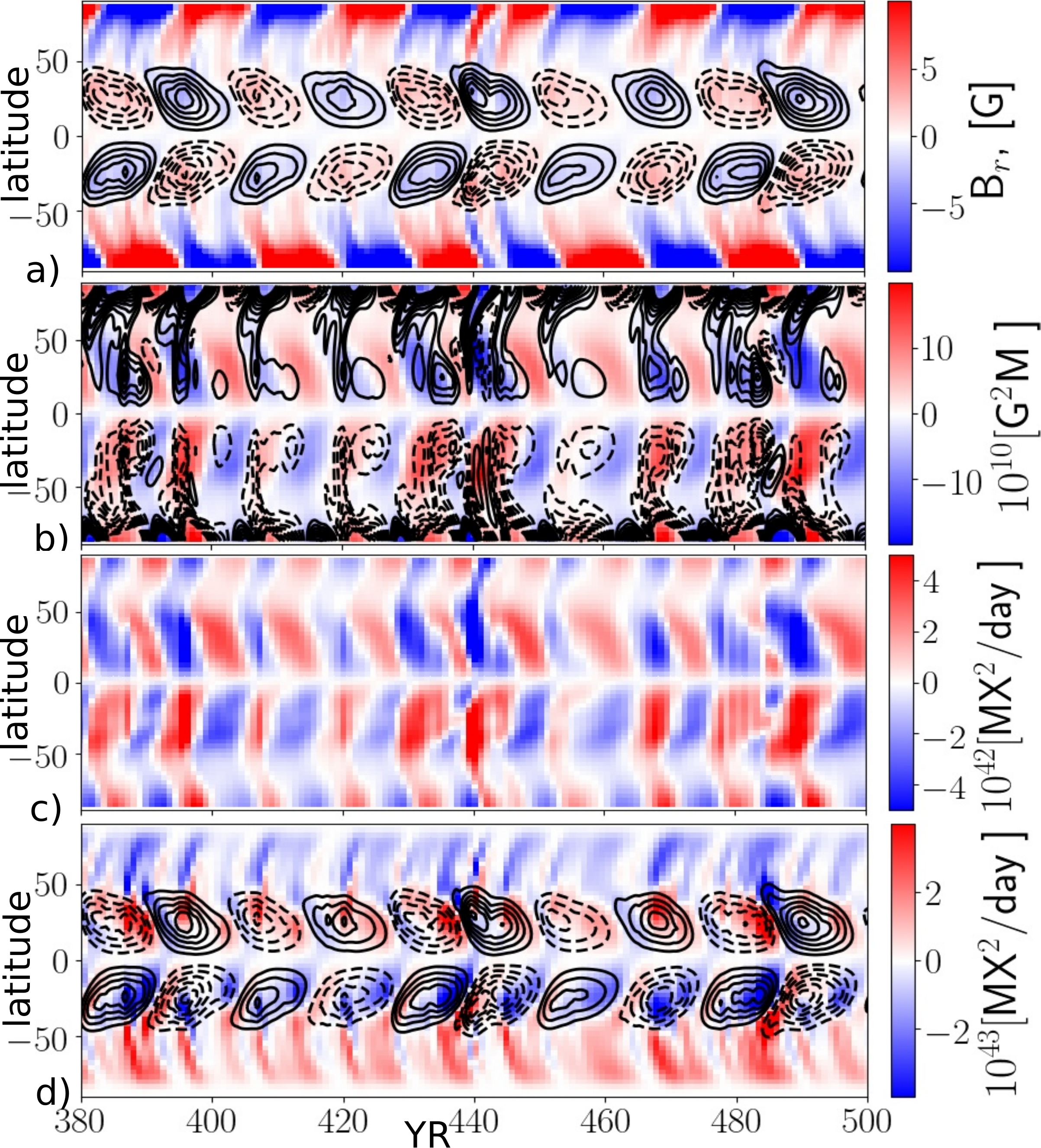}\caption{\label{fig-tl-1}The same as Figure \ref{fig-tl} for the model M3.}
\end{figure}

{The Figures \ref{fig-tl} and \ref{fig-tl-1} illustrate
the time-latitude diagram of the toroidal magnetic field in the upper
part of the solar convection zone, the radial magnetic field at the
surface, the time-latitude evolution of the small-scale magnetic helicity
density and the fluxes $F_{L}$, $F_{S}$ . 
The series include episodes of the high and relatively weak magnetic activity for the models M2
and M3. 
The similar evolution diagrams were found for the models M1 and M4.  Evolution patterns of the magnetic field and the HHR for the
large- and small-scale magnetic helicity density are qualitatively
similar to our previous results which were discussed in \citet{Pipin2013d}.
In particular, for the episodes of high activity, e.g., during years 380 - 400, our models show the standard HHR for the small- and the large-scale
magnetic fields. 
he inversions of the HHR occur during the relatively short periods of the growing and decaying phases of the magnetic cycles.
In the weak cycles these episodes last a longer time. 
The similar tendency there is in the hemispheric behavior of the helicity fluxes.
The results for the helicity flux due to the differential rotation, i.e., $F_{\Omega}$, are qualitatively similar to the surface flux transport simulations of \citet{Hawkes2019AA}. 
The amplitude of this flux in our case is by an order of magnitude smaller than in theirs' because we are restricted to the contribution of the axisymmetric magnetic field. In our simulations the dynamo regimes show that $F_{L}<F_{S}<F_{\Omega}$. }

{It is found that in the given range variations of the parameters,
$\xi^{(\chi)}$, $\xi^{(m)},$ the variations of the dynamo efficiency are weak. 
The magnitude of the maximum total energy of the generated
magnetic field among the models varies about 20\% (see the Table 1).
This is likely because the maximum of the dynamo wave is located near the bottom of the convection zone and the dynamo efficiency does not change much because the helicity flux from the surface. 
On the other hand, it is assumed that random variations of the $\alpha$ -effect are uniform in radius. 
Therefore, its effect is more profound than the effect of the helicity flux variations.}

\begin{figure}
\centering \includegraphics[width=0.95\textwidth]{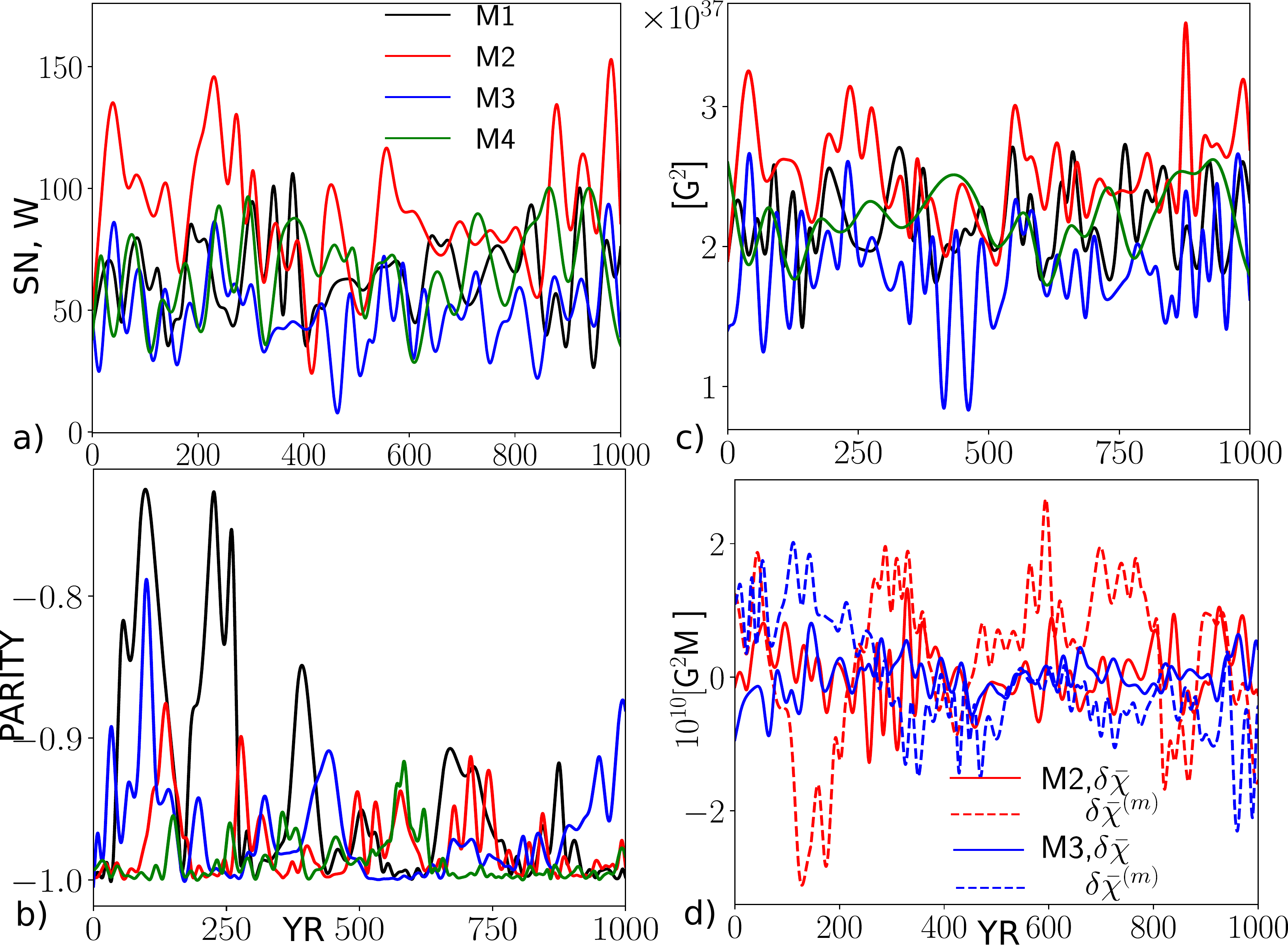}\caption{\label{fig:imbalance}(a) The simulated value of the sunspot number; (b) the parity index; (c) the net magnetic helicity density at the surface}
\end{figure}

Figure \ref{fig:imbalance} shows the simulated value of the sunspot
number, $W$, the parity index, $P$ and the net magnetic helicity
density at the surface for the small-scales and the large-scale magnetic fields, $\delta\overline{\chi}$ and $\delta\overline{\chi}^{(m)}$,
respectively, {for our models. 
In this Figure we filter out all variations with periods smaller than 20 years. }

\begin{figure}
\includegraphics[width=0.9\textwidth]{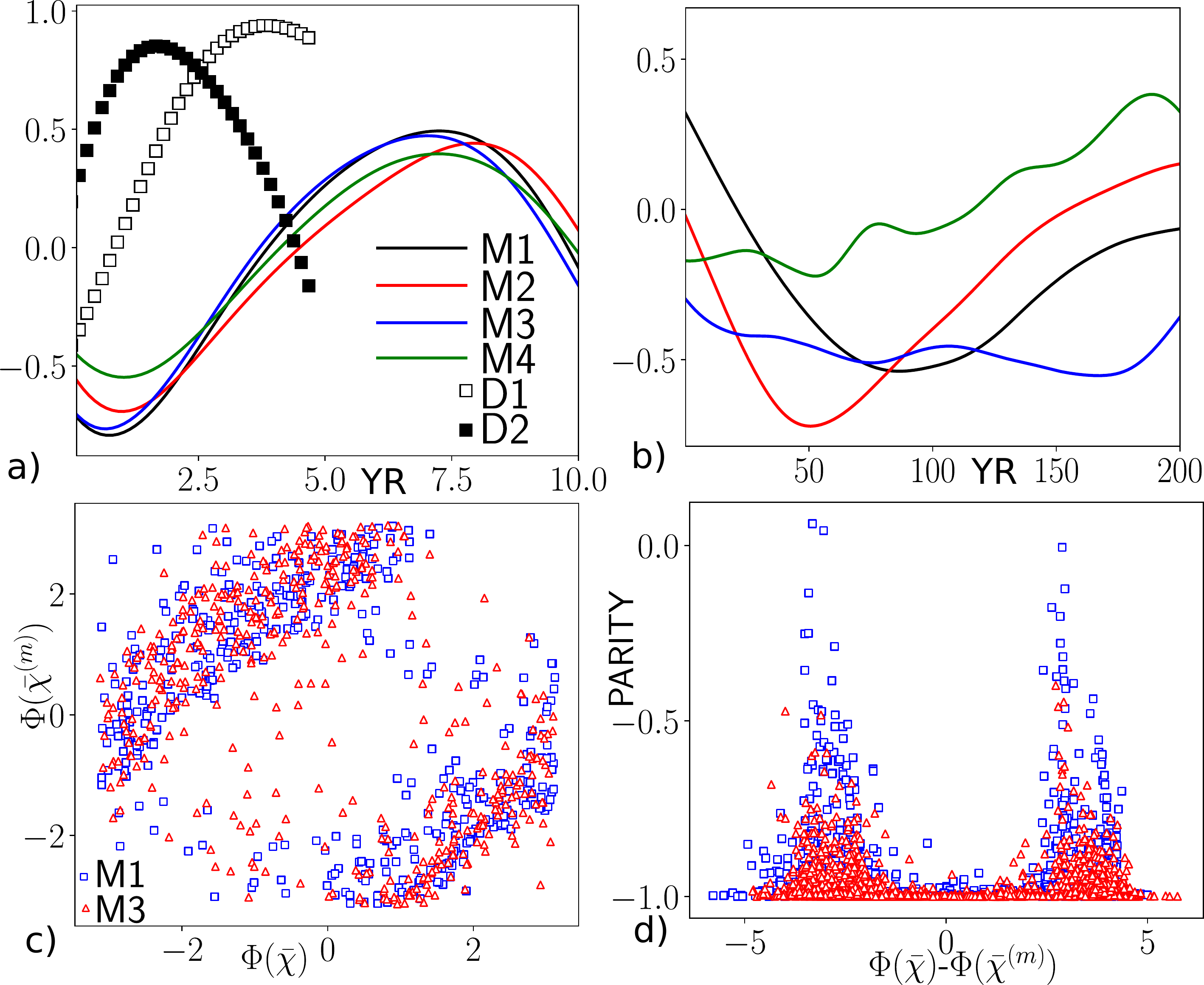}\caption{\label{fig:output}{(a) The cross-correlations $\left\langle \delta\overline{\chi}\left(t\right)\delta\overline{\chi}^{(m)}\left(t+\tau\right)\right\rangle $
for the dynamo models and the $\left\langle \delta\overline{h}_{C}(t)\delta\overline{\chi}^{\left(m\right)}\left(t+\tau\right)\right\rangle $
from the observational data sets of of $h_{C}$ and $\overline{\chi}^{(m)}$
(D1) and the same correlations computed from the first first 3 empirical modes (D2); 
(b) the same as (a) calculated from the smoothed series
of the model runs see Figure \ref{fig:imbalance}; (c) the phase diagram for phases of the analytical signals of the $\delta\overline{\chi}\left(t\right)$
and $\delta\overline{\chi}^{(m)}\left(t\right)$ in the models M1
and M3; 
(d) the parity index in the models M1 and M3 vs the difference
of the phases of the analytical signals of the $\delta\overline{\chi}\left(t\right)$
and $\delta\overline{\chi}^{(m)}\left(t\right)$.}}
 
\end{figure}

{The results of the model M3 show that during epochs of the centennial magnetic activity minima, corresponding to periods around 100, 400, and 900 years, the distributions of the magnetic field and the large-scale magnetic helicity density are not symmetric about equator and the parity index is greater than -1 during the most part of the cycle, oscillating around $[-0.8:-0.6]$. 
The model M2 shows the strongest
variations of the sunspot number parameter, W, among the models. In this model high variations of the magnetic parity and the helicity density imbalances are, sometimes, found for the episodes of the centennial maxima. 
In these models the increase of the parity index seems to
be accompanied and connected with the increase of the oscillation
magnitude of the imbalances $\delta\overline{\chi}$ and $\delta\overline{\chi}^{(m)}$.
Variations of these parameters on the short time scale including those within the range of 1 year go in anti-phase. }
This effect is quantified by the Pearson correlation coefficient $\left\langle \delta\overline{\chi}\left(t\right)\delta\overline{\chi}^{(m)}\left(t+\tau\right)\right\rangle $.
It is further illustrated by Figure \ref{fig:output}(a), where we
show results for{ $\left\langle \delta\overline{h}_{C}(t)\delta\overline{\chi}^{\left(m\right)}\left(t+\tau\right)\right\rangle $
computed from our observational data sets} and from the models using the original data sets and the smoothed ones where {we filter out all variations with periods smaller than 20 years.}

{We see that in all the models there is an anti-correlation between the $\delta\overline{\chi}\left(t\right)$ and $\delta\overline{\chi}^{(m)}\left(t+\tau\right)$
if $\tau=0$. The effect of the anti-phase synchronization is largest in case of the models M1 and M3. 
The same effect is present in the data as well, in particular, when we restrict ourselves with the first
three empirical modes of $h_{C}$ and $\overline{\chi}^{(m)}$ (the
curve D2). 
This conclusion is not robust because the correlation coefficient changes
sign to positive after adding the forth empirical mode (see the curve D1). Also, we have to take into account that quality of the  observational data is not sufficient for the
robust conclusion. The anti-phase synchronization $\delta\overline{\chi}\left(t\right)$
and $\delta\overline{\chi}^{(m)}\left(t\right)$ persists in the model M3 over the centennial time scales as well.  
Oppositely, the model M4
show synchronization on the long -time scales. The results of the
models M1 and M2 show the anti-phase behavior over the range of scales
from 50 to 100 years and they show the absence of the significant
correlations on the longer time intervals. }The anti-phase synchronization
in the models M1 and M3 is further illustrated by the phases of the
analytical signals of $\delta\overline{\chi}\left(t\right)$ and $\delta\overline{\chi}^{(m)}\left(t\right)$,
which are computed using the Hilbert transform, and denoted as $\Phi\left(\overline{\chi}\right)$
and $\Phi\left(\overline{\chi}^{(m)}\right)$, respectively. In the
model M1 the synchronization persists on the longer time intervals
than in the model M3. This likely due to absence of the forced magnetic
helicity fluxes. The effect of synchronization is reflected in the
clustering of the points in the phase diagram to the two bands. The
effect is less for the model M3. The relation of the synchronization
between the $\delta\overline{\chi}\left(t\right)$ and $\delta\overline{\chi}^{(m)}\left(t\right)$
with the magnetic parity is further illustrated in Figure \ref{fig:output}(bottom).
We see that the dispersion of the difference $\Phi\left(\overline{\chi}\right)-\Phi\left(\overline{\chi}^{(m)}\right)$
is large (and possibly random nature) when the parity index varies
around 1. The dispersion decreases when the parity index grows.

\section{Discussion and Summary}

Results of our models predict the anti-correlation between variations of magnetic helicity imbalance on the small- and large-scales on the short time intervals. The similar effect is demonstrated by the observational data (see, Figure \ref{fig:imb}). However, observations are rather noisy and cover a small period of time which is not enough to robustly determine the given effect. It is found that the hemispheric asymmetry of the magnetic helicity flux can affect the hemispheric asymmetry of the magnetic activity. The latter is characterized the parity parameter P (see,  Eq(\ref{eq:parity})), and the helicity imbalance parameters. The parity parameter of the
dynamo generated magnetic field is related with mixing of the fundamental
dynamo modes corresponding to the symmetric and antisymmetric about
the solar equator magnetic field \citet{Sokoloff1994}.

The net magnetic helicity density of the large-scale magnetic field
can be related with the parity parameter $P$ as well. To see it,
lets decompose $r$ and $\phi$ components of the magnetic field and
its vector potential on series of the Legendre polynomial $P_{n}$
and $P_{n}^{1}$ (also see \citet{Pipin2014b}): 
\begin{equation}
\bar{A}_{\phi}\left(t,\theta\right)=\sum_{n=1}^{N}a_{\phi}^{(n)}\left(t\right)P_{n}^{1}\left(\cos\theta\right),\label{eq:aa}
\end{equation}
\begin{equation}
\bar{B}_{r}\left(t,\theta\right)=\sum_{n=1}^{N}b_{r}^{(n)}\left(t\right)P_{n}\left(\cos\theta\right),\label{eq:br}
\end{equation}
\begin{equation}
\bar{B}_{\phi}\left(t,\theta\right)=\sum_{n=1}^{N}b_{\phi}^{(n)}\left(t\right)P_{n}^{1}\left(\cos\theta\right),\label{eq:bf}
\end{equation}
\begin{equation}
\bar{A}_{r}\left(t,\theta\right)=\sum_{n=1}^{N}a_{r}^{(n)}\left(t\right)P_{n}\left(\cos\theta\right).\label{eq:arf}
\end{equation}
Using the standard relations between $P_{n}$ and $P_{n}^{1}$ we
can find expressions for coefficients of the vector-potential: 
\begin{eqnarray}
a_{\phi}^{(n)}\left(t\right) & = & -\frac{Rb_{r}^{(n)}\left(t\right)}{n\left(n+1\right)},\label{eq:nn}\\
a_{r}^{(n)}\left(t\right) & = & -Rb_{\phi}^{(n)}\left(t\right)\label{eq:nnb}
\end{eqnarray}
Then restricting ourselves only with two first terms of expansions
we have 
\begin{eqnarray*}
\bar{A}_{\phi}\left(t,\theta\right) & = & -\frac{R_{\odot}b_{r}^{(1)}\left(t\right)}{2}P_{1}^{1}-\frac{R_{\odot}b_{r}^{(2)}\left(t\right)}{6}P_{2}^{1}+\dots\\
\bar{B}_{\phi}\left(t,\theta\right) & = & b_{\phi}^{(1)}\left(t\right)P_{1}^{1}+b_{\phi}^{(2)}\left(t\right)P_{2}^{1}+\dots
\end{eqnarray*}

Note that, $\int_{-1}^{1}\bar{A}_{\phi}\bar{B}_{\phi}d\mu=\int_{-1}^{1}\bar{A}_{r}\bar{B}_{r}d\mu$
because of the symmetry properties \citep{Blackman2003}. Therefore,
the net magnetic helicity density will be 
\begin{equation}
\delta\overline{\chi}^{(m)}=2\int_{-1}^{1}\bar{A}_{\phi}\bar{B}_{\phi}d\mu\approx-\frac{4R_{\odot}}{3}b_{r}^{(1)}\left(t\right)b_{\phi}^{(1)}\left(t\right)-\frac{8R_{\odot}}{15}b_{r}^{(2)}\left(t\right)b_{\phi}^{(2)}\left(t\right)+\dots
\end{equation}
In this Equation, $b_{r}^{(1)}$ is the dipole mode of the radial
magnetic field and $b_{\phi}^{(1)}$ is the quadrupole mode of the
toroidal magnetic field. Therefore, the magnetic parity $P$ is readily
related with magnetic helicity imbalance. In the recent paper of \citet{Pipin2018e},
the parameter $P$ was calculated from the data set including the
last 4 solar cycles. It was found that $P\approx0$ (strong asymmetry
of the magnetic activity) near the maxima of the sunspot activity.
Taking into account the data presented in Figure \ref{fig:imb} we
conclude that the models prediction about the connection of the parity
and helicity imbalance parameter roughly agrees with observations.
{On the other hand, in the long-term variations there may exist
a correlation of magnetic helicity imbalance on the small- and large-scales.
The specific situation depends on the nature of the helicity fluxes.
The mix of the short-time random fluxes $F_{L}$ and $F_{S}$ results
to correlation of the $\overline{\chi}$ and $\overline{\chi}^{(m)}$
on the long-term intervals more than 100 year. 
While the predominance of one of the $F_{L}$ or $F_{S}$ can result in anti-correlation over this time intervals. }

{In our dynamo model set the magnitude of the helicity flux
variations is much less than the helicity flux due to differential
rotation. Our results show that the problem of the magnetic helicity
flux from the dynamo domain can be a delicate question. The helicity
flux due to differential rotation is determined by the top boundary
condition. 
In our model we use the boundary  conditions that provide penetration of 
the toroidal magnetic field into the surface and the potential poloidal magnetic  outside the dynamo domain (see \citealt{Pipin2019c}). 
The effect of such condition on the helicity conservation is not well studied.
The results of this paper show that the relatively small magnitude of the helicity fluxes from the subsurface of the Sun can affect the dynamo evolution.} 

The predicted patterns of the small- and large-scale magnetic
helicity in our dynamo model are in qualitative agreement with results
of observations of the current helicity of solar active regions (see,
\citealt{Zhang2010}) and results of \citet{Pipin2019a}. {The time-latitude
evolution of the helicity fluxes $F_{S}$,$F_{L}$ are similar to
those shown by , $\overline{\chi}$ and $\overline{\chi}^{(m)}$ which
is expected by the model design. The helicity flux due to the differential
rotation, $F_{\Omega}$, evolves a bit differently, and its evolution
in our models is in agreement with results of \citet{Hawkes2019AA}
(cf., our Fig3a and Fig3d in their paper). The interesting feature
of the helicity flux found in both papers is the presence of both
signs of the $F_{\Omega}$ simultaneously as the dynamo cycle progress
from high to low latitudes. The equatorial part of diagrams satisfy
the standard HHR. In our models the given effect can be explained
by the magnetic cycles overlap. This effect was discussed recently
by \citet{Pipin2019c} and this discussion can be extended to results
of our paper as well. It is interesting that a rather similar pattern
can be found in the nonaxisymmetric dynamo model of \citet{Pipin2018e}.
\citet{Pipin2019a} used this model as a benchmark prior to processing the magnetic vector synoptic maps of SDO/HMI. }

For the given parameters of the helicity flux variations the amplitude
of the dynamo wave does not change strongly. The most effect is found
for the magnetic field equatorial symmetry and the magnetic helicity
imbalance variations. According to dynamo theory \citep{Blackman2003},
the magnetic helicity of large-scale field , is, in general but not
completely, determined by the sign of $\alpha$-effect and the opposite
helicity sign is expected for the small-scale magnetic field. A complicated
connection between small- and large-scale properties of the magnetic
helicity fluxes in solar cycles 23-24 was discussed earlier by \citet{Yang2012}
and \citet{Zhang2013}. Results of \citet{Brandenburg2017}, \citet{Singh2018}
and \citet{Pipin2019a} show that the bi-helical property can be violated
and it was violated in solar cycle 24. As a result, the sign of the surface magnetic helicity density of the large- and small-scale field
can be the same. 
The origin of this phenomenon is unclear. {In
particular, the results of \citet{Brandenburg2017} of mono-helical magnetic helicity spectrum was shown to become bi-helical by \citet{Singh2018}
when data from SOLIS was used instead of HMI. 
Also, \citet{Singh2018}
found almost always bi-helical spectra, mono-helical ones being clearly very rare. 
In a qualitative agreement with results of \citet{Pipin2013d}
and results of this paper, \citet{Singh2018} found that the sign
rule in between the large and small scale helicities can reverse,
especially during the declining and minimum phases.} In general,
we can assume its relationship with fluctuations of magnetic helicity fluxes. Our results about anti-correlation between variations of magnetic helicity imbalance on the small- and large-scales support this idea.
With some reservation, it can be suggested that there is a relationship between violation of the bi-helical property on the surface and the equatorial parity breaking of the magnetic activity evolution. In this study we show the theoretical possibility of such a relation.
However, the strength of our prediction is rather limited because
the amplitude of the helicity flux fluctuations remains unconstrained in the model. 
This opens an interesting theoretical and observational
prospects for the future studies.

\textbf{Acknowledgements} 
The Authors would like to thank the anonymous Referees for their constructive criticism and comments
This study is supported by the RFBR of Russia
and NNSF of China bilateral grant 19-52-53045, also grants 11427901, 10921303, 11673033, U1731113,11611530679, and 11573037 of the National Natural Science Foundation of China; by grant No. XDB09040200, XDA15010700 of the Strategic Priority Research Program of Chinese Academy of Sciences;
by the Max-Planck Society Inter-Institutional Research Initiative
Turbulent transport and ion heating, reconnection, and electron acceleration in solar and fusion plasma of Project No. MIF-IF-A-AERO8047; by Max-Planck-Princeton Center for Plasma Physics PS AERO 8003; by ISSI International Team on Magnetic Helicity estimations in models and observations of the solar magnetic field. The authors would also like to thank the Supercomputer Center of the Chinese Academy of Sciences (SCCAS) and the Max-Planck Computing and Data Facility (MPCDF) for the allocation of computing time. 
The dynamo model was tested during ``Solar Helicities in Theory
and Observations: Implications for Space Weather and Dynamo Theory''
Program at Nordic Institute for Theoretical Physics (NORDITA). DDS
would like to acknowledge support from the RFBR grant 18-02-00085. VVP
has run dynamo simulations presented in the paper using the computer
cluster facilities of ISTP and conducted this work as a part of research project II.16.3.1 of ISTP SB RAS. 

\bibliographystyle{jpp}

\end{document}